\documentclass{iopart}

\usepackage{fleqn}

\usepackage{ifthen}
\usepackage{ifpdf}

\usepackage{latexsym}
\usepackage{amssymb}
\usepackage{bm}

\ifpdf
\usepackage{graphicx}
\usepackage{epstopdf}
\else
\usepackage{graphicx}
\usepackage{epsfig}
\fi

\newcommand{\trc}{\mbox{trace}}

\newcommand{\tbox}[1]{\mbox{\tiny #1}}

\newcommand{\be}[1]{\begin{eqnarray}\ifthenelse{#1=-1}{\nonumber}{\ifthenelse{#1=0}{}{\label{e#1}}}}
\newcommand{\ee}{\end{eqnarray}}

\newcommand{\hide}[1]{}

\newcommand{\putgrf}[2][\hsize]{ \includegraphics[width=#1]{#2} }

\begin{document}


\title[Multimode Conductance]
{The Multimode Conductance Formula \\ for a Closed Ring}

\author{Doron Cohen and Yoav Etzioni}

\address{
Department of Physics, Ben-Gurion University, Beer-Sheva 84105, Israel
}


\begin{abstract}
The multimode conductance of a {\em closed} ring is found   
within the framework of a scattering approach. The expression 
can be regarded as a generalization of the Landauer formula. 
The treatment is essentially {\em classical} because we assume 
short coherence time. Our starting point is the Kubo formalism, 
but we also use a master equation approach for the derivation. 
As an example we calculate the conductance of a multimode 
waveguide with an attached cavity. 
\end{abstract}

\section{Introduction}

The notion of conductance has gone several 
transformation in the last century.
In the mesoscopic community \cite{datta,imry}, 
following Landauer, it is customary nowadays 
to consider the open geometry that is described 
in Fig.1a, where a device is attached to left 
and right reservoirs, and the bias is 
understood as emerging from a chemical potential 
difference. For a single mode device it is 
argued that the conductance is essentially 
the transmission, while for a multimode device 
with ``spinless" electrons
\be{1}
G_{\tbox{Landauer}}
= \frac{e^2}{2\pi\hbar}\sum_{n,m} g^T_{nm}
\ee  
where $g^T_{nm}$ is the transmission matrix. 
We can optionally assume that 
the chemical potential of the two reservoirs 
is the same, and consider the effect of 
an electro motive force (EMF) such that the 
voltage drop is concentrated across a segment 
of the device. More generally we can 
consider the problem of driving a current 
by changing the electrical potential at some region 
of the device. The latter more general problem 
is known as ``quantum pumping''. The calculation 
of the ``geometric" conductance in the latter case 
leads to the B\"{u}ttiker-Pr\'{e}tre-Thomas (BPT) formula \cite{BPT}, 
which is a generalization of the Landauer formula Eq.(\ref{e1}).

\begin{figure}[t]
\putgrf[0.9\hsize]{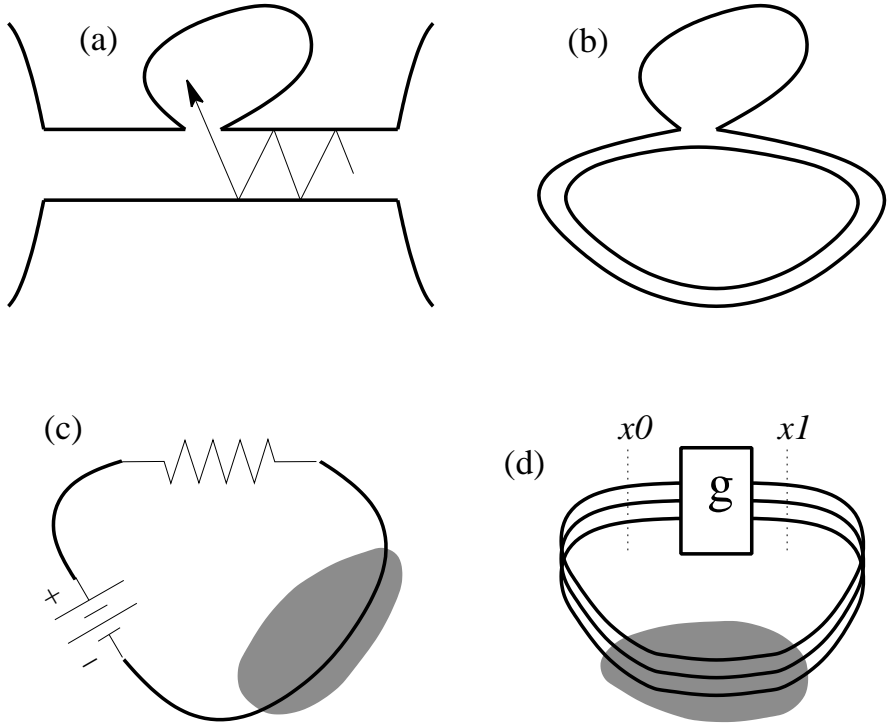} 
\caption{Panel (a) displays the standard Landauer 
(open) geometry of connecting a conductor to a left and a right 
reservoirs. In this illustration the conductor is a rectangular 
waveguide to which a cavity is attached. In panel (b) the leads 
are joined together so as to form a ring. The motion in the ring 
is assumed to be chaotic due to the scattering in and out of 
the cavity. Panel (c) is the schematic electrical engineering 
representation of the system. In panel (d) the system is 
modeled as a network. The scattering region is described 
by the transition matrix $g_{ab}$. 
In (c) and in (d) we indicate the presence of the environment 
by the gray shading. In fact (a) can be regarded as a special 
case of (b) provided one assumes that the 
effect of the environment is to randomize the velocity 
within the wire region. The current is measured 
via the section $x=x_1$. The Electro motive force (EMF) 
is realized by time dependent Aharonov Bohm flux. 
The voltage drop can be concentrated anywhere along 
the ring (say across $x=x_0$). 
Setting a chemical potential difference in the setup 
of panel (a) can be regarded as such particular option.}
\end{figure}

It is quite natural to ask what happens 
if the two leads are detached from 
the reservoirs, and the system is closed
into a ring as in Fig.1b \cite{pmo,pme}. 
Still we can induce EMF by changing 
in time an Aharonov-Bohm magnetic flux, 
or we can change the potential in some 
region of the device, so as to get 
an electrical current. 
In spite of much interest in closed mesoscopic rings 
a straightforward answer to this simple 
question has not been given. We shall 
review later the main published statements 
\cite{rings,debye,IS,locl1,gefen,orsay}

In this paper we are interested in circumstances 
such that the leading result for the conductance 
is of classical nature. This is completely 
analogous to the discussion of {\em diffusive} rings   
in circumstances such that the leading result is 
given by the Drude formula. We are going to assume 
that the coherence time is shorter than the time 
that it takes for an electron to encircle the ring. 
Thus, as far as the dynamics is concerned, our 
analysis is essentially of {\em classical} nature.  
{\em The case of a fully coherent ring requires 
further analysis which is beyond the scope 
of the present work}. Therefore it  
will be discussed in separate publications: 
The one-mode case in \cite{kbr} 
and the multi-mode case in \cite{kbw}.

It is important to realize that 
our assumption of short coherence 
time is plays analogous role 
to that which is played by the 
reservoirs in the Landauer-BPT formalism. 
In fact we are going to explain that
the Landauer-BPT formalism can be regarded 
as a {\em special} limit of the problem that 
we are going to consider. 
This will further illuminate the formal 
discussion in \cite{pmo}.

\subsection{Motivation}

The interest in the response of small mesoscopic 
rings is long standing \cite{rings,debye,IS,locl1,gefen}.
Measurements of the the conductance of closed 
mesoscopic rings has been performed already~10 years 
ago \cite{orsay}. In a practical experiment 
a large array of two dimensional rings is fabricated.
The conductance measurement can be achieved  
via coupling to a highly sensitive electromagnetic 
superconducting micro-resonator. In such setup the EMF 
is realized by creating a current through 
a "wire" that spirals on top of the array,  
and the conductance of the rings is determined via 
their influence on the electrical circuit. 
Another possibility is to extract the conductance  
from the rate of Joule heating. The later can 
be deduced from a temperature difference measurement 
assuming that the thermal conductance is known.

We do not think that there is any problem to design 
rings of the type which is illustrated in Fig.~1.       
The question is why to bother? This brings us to the 
theoretical motivation for this work. Past theoretical 
studies are quite well summarized by the paper 
``(Almost) everything you always wanted to know about 
the conductance of mesoscopic systems" \cite{gefen} 
and see references therein. The major interest was 
in {\em diffusive rings}, and the main issue 
was {\em weak localization} corrections to the 
{\em Drude result} taking into account 
the level statistics, the type of occupation, etc.   
Our interest is in a different type of configuration,   
which is motivated by ``quantum chaos" studies. 
Our claim (see next subsection) that the conductance 
of a closed ring can be larger than the number 
of open modes is quite {\em unorthodox}.

More importantly: Our approach to the problem 
of mesoscopic conductance has a {\em practical appeal}.
The reason for the popularity of the scattering 
approach in mesoscopic physics  
is its "plug and play" feature. The experimentalist 
is able to characterize the scattering properties 
of his/her device, and then he/she is able to make 
a prediction regarding the conductance. 
It is only natural to extend this "plug and play" 
approach to the analysis of conductance of {\em closed} rings.      
This extension is far from being trivial, 
as we explain later (in section~6).

We care to make clear distinction
between classical and quantal effects. 
This helps to develop a better intuition for 
the physics of such devices. 
The approach in the present work 
is in the spirit of the Boltzmann picture: 
The scattering ``cross section", which is 
possibly of quantum mechanical nature, 
is taken as an input, while the overall dynamics 
is assumed to be of classical nature.
The implications of quantum interference will be discussed  
in future works \cite{kbr,kbw}.

\subsection{specific results}

The conductance of a single mode ring ($\mathcal{M}=1$) 
with a stochastic scatterer that has transmission $g_T$ 
is given by the expression 
\be{2}
G=\frac{e^2}{2\pi\hbar}\left[\frac{g_T}{1-g_T}\right]
\ee  
We shall show that the multimode generalization of this 
formula is:
\be{3}
G = 
\frac{e^2}{2\pi\hbar} 
\sum_{nm} \left[ 2g^T/(1-g^T+g^R) \right]_{nm}
\ee  
where $g^T_{nm}$ and $g^R_{nm}$ are the 
$\mathcal{M}\times\mathcal{M}$  transmission 
and reflection blocks of the transition matrix, 
%
%
%
As an example we analyze the system of Fig.1a.  
We find that the Landauer formula gives 
\be{4}
G_{\tbox{Landauer}} 
\approx  \frac{e^2}{2\pi\hbar}
\left[ 1 - \frac{1}{4}\alpha \right] 
\mathcal{M}
\ee
where $\alpha$ is the dimensionless size 
of the opening to the cavity. In contrast to that 
for the multi-mode conductance of the 
corresponding ring structure (Fig.1b) we get  
\be{5}
G \approx 
\frac{e^2}{2\pi\hbar}\left[ 
\frac{1}{\alpha} (1+\ln(2\mathcal{M})) 
\right] 
\mathcal{M}
\ee
Unlike the case of the Landauer conductance, the 
result does not reflect the number of open modes.  
This is because the contribution of the low modes 
is singular in the limit of small $\alpha$. Furthermore, 
the conductivity (conductance per channel) diverges 
logarithmically in the classical limit.

\subsection{Outline}

We define the model system in sections~2 and~3  
and the notion of conductance in section~4.
In section~5 we argue that the notion of 
conductance is meaningful even in the 
absence of a bath.  We further discuss 
the role of the environment in section~6, 
where we make distinguish between 
various type of bath induced effects.  

As explained in section~7 the purpose 
of the ``linear response analysis" 
is to find the stationary-like state 
of the driven system.  
The procedure is to assume that 
in the absence of driving the system 
would be in a (strict) stationary state, 
which we regard as the zero order solution. 
Then we try to find a first order solution (in the EMF)  
to the time dependent problem.  

In sections~8 and~9 we discuss 
the Kubo approach to linear response.
We  take the simplest route following 
Refs.\cite{wilk,jar,frc,pmc} leading to the 
fluctuation-dissipation version of the Kubo formula.
The application of the Kubo formula 
to the analysis of the single mode 
conductance is presented in sections~10 and~11. 
The relation to the Landauer result 
is clarified in section~12.   

There are many (equivalent) ways 
to do ``linear response analysis". 
It turns out that the derivation of 
the multi-mode conductance formula 
becomes more transparent by 
adopting a master equation approach.
This is carried out in section~13. 

Throughout the paper, and in particular 
in the concluding section, we emphasize 
that our approach has a straightforward 
extension to the analysis of quantum pumping. 
The Kubo approach allows a better 
understanding of the role which is 
played by the environment, and makes it 
possible to bridge between the strict 
quantum adiabatic limit and the other 
extreme limit of having an open geometry.

\section{Setting up the model}

Consider non-interacting spinless electrons in a ring, 
as in Fig.1b. The one-particle Hamiltonian is 
\be{6}
{\cal H} = \frac{1}{\mathsf{m}} (\bm{p} - e\Phi a(\bm{r}))^2 + V(\bm{r})
\ee
where $\mathsf{m}$ and $e$ are the mass 
and the charge respectively. 
The vector potential which is associated 
with the flux $\Phi$  is described by      
\be{7}
\oint a(\bm{r}) \cdot d\bm{r} = 1
\ee
The dimensionality of the ring is $d$. 
The ring consists of a ``wire" region and 
a scattering region.
The motion of the particle 
inside the ring is assumed to be globally 
chaotic. The coordinate along the wire will be 
denoted as $x$. The scattering region 
is located at $x \sim 0$.

In the geometry of Fig.1b the ``wire" 
is a $d=2$ waveguide of width $L_{\perp}$. 
Later we describe the waveguide 
as a set of $d=1$ wires (Fig.1d)  
such that each ``wire" corresponds 
to an open mode of the waveguide.   
The length of the ring $L$ 
is assumed to be large compared with the 
scattering region so as to allow 
meaningful definition of a scattering matrix 
in the quantum mechanical analysis 
(evanescent modes are ignored).

The ring is driven by a time dependent 
Aharonov-Bohm flux. The EMF $-\dot{\Phi}$ is assumed 
to be constant. There are various ways to introduce 
the EMF into the ring. One possibility is 
to have all the voltage drop over 
a section at $x=x_0$. Namely, 
\be{8}  
a(x)=\delta(x-x_0)
\ee
For sake of later analysis we define a generalized 
force which is associated with the flux:
\be{9}
{\cal F} = -\frac{\partial {\cal H}}{\partial \Phi} 
= e\hat{v} \delta(\hat{x}-x_0)
\ee
where $v$ is the velocity in the $x$ direction. 
In the quantum mechanical case a symmetrization is implicit.
This is in fact a current operator. 
Obviously we do not have to measure 
the current at the same point where 
we apply the voltage. So for sake 
of generality we introduce the notation 
\be{10}
\mathcal{I} = e\hat{v} \delta(\hat{x}-x_1)
\ee
We also note that with uniform averaging over $x_1$ 
we get $(e/L)\hat{v}$ which is 
essentially the velocity operator.

In the absence of driving 
the ``pure" stationary states of the system are 
the microcanonical states. We use classical language 
but also have in mind a semiclassical picture. 
Each microcanonical state occupy a shell 
whose phase space volume is $(2\pi\hbar)^d$. 
The density of states is 
\be{11} 
\mathsf{g}(E)= \int\int \frac{d\mathbf{r} d\mathbf{p}}{(2\pi\hbar)^d} 
\delta(E-\mathcal{H}(\mathbf{r},\mathbf{p}))
\ee
The zero order stationary state is characterized 
by an occupation function $f(E)$. Later we shall take 
it to be the Fermi function. Thus
\be{12}
dN 
= \rho(\mathbf{r},\mathbf{p}) 
\frac{d\mathbf{r} d\mathbf{p}}{(2\pi\hbar)^d} 
= f(E)  
\frac{d\mathbf{r} d\mathbf{p}}{(2\pi\hbar)^d} 
\ee
where $E=H(\bm{r},\bm{p})$ is the energy.
The distribution of the particles in energy is 
\be{13}
\rho(E)=\mathsf{g}(E)f(E)
\ee
and the total number of particles is 
\be{14}
N = \int\int \rho(\mathbf{r},\mathbf{p})  
\frac{d\mathbf{r} d\mathbf{p}}{(2\pi\hbar)^d}
= \int_{-\infty}^{\infty} \rho(E) dE 
= \int_{-\infty}^{\infty} f(E) \mathsf{g}(E) dE   
\ee

\section{Network modeling}

A network is defined as a set of 1D wires 
that are connected in vertices. 
The network Hamiltonian is ill defined 
in the classical limit because the 
the scattering in each vertex  
is described by a scattering matrix.
In particular for the model of Fig.1d 
the scattering is described by 
a scattering matrix $S_{ab}$, 
and we define the corresponding 
transition matrix as $g_{ab}=|S_{ab}|^2$.
Thus the classical description 
of the system is stochastic 
rather than deterministic.

Still we can regard networks as an effective 
way to describe the chaotic dynamics \cite{kottos}.
The reason is that upon coarse 
graining a chaotic system looks like 
a stochastic model. 
Specifically, in the case of the system of Fig1b, 
quantum mechanics introduces 
``coarse graining" in a most natural way. 
Each mode in the scattering  
problem can be regarded as a 1D wire 
with the dispersion relation
\be{15}
v_n = \frac{1}{\mathsf{m}} \sqrt{2\mathsf{m}E 
- \left(\frac{\pi\hbar}{L_{\perp}}n \right)^2}
\ee
where $E$ is the energy of the particle, 
and $L_{\perp}$ is the width of the waveguide.  
The open modes are are those 
for which $v_n$ is a real number. 
We shall denote the number 
of open modes by $\mathcal{M}$ 
hence the number of open channels 
in the open geometry is $2\mathcal{M}$.

The density of states of the system 
can be written as a sum over single-mode 
expressions:   
\be{16}
\mathsf{g}(E) =  \mathsf{g}_{\tbox{dot}}(E) 
+ 2\sum_{n=1}^\mathcal{M} \frac{L}{(2\pi\hbar)} \frac{1}{v_n(E)}
\ee
where the factor of two takes into account 
both clockwise and anticlockwise motion.
A stationary state of the system is described 
by the distribution functions $\rho_n^{\rightarrow}(E)$ 
of the clockwise moving particles and $\rho_n^{\leftarrow}(E)$ 
of the counter-clockwise moving particles. The index $n$ 
distinguishes different modes. The normalization is such that 
\be{17}
N = \int \rho_{\tbox{dot}}(E) dE + L\sum_n
\int (\rho_n^{\rightarrow}(E)+\rho_n^{\leftarrow}(E)) dE 
\ee
The density of particles per unit length 
in a given mode is implied by Eq.(\ref{e13}):  
\be{18}
\rho_n^{\rightarrow}(E)dE 
= \rho_n^{\leftarrow}(E)dE 
= \frac{dE}{(2\pi\hbar) v_n(E)} f(E)
\ee
Note that for a microcanonical 
distribution $dE$ can be regarded 
as a fixed parameter that defines  
an energy window or a width 
of an energy shell.

The scattering is described by 
a $2\mathcal{M} \times 2\mathcal{M}$  
transition matrix that has a block structure:  
\be{19}
g_{ab} = \left( \matrix{g^R & g^T \cr g^T & g^R}  \right) 
\ee  
It consists of the reflection matrix $g^R_{nm}$
and the transmission matrix $g^T_{nm}$. 
Note that the channel index $a$ 
contains both mode specification and  
left/right lead specification.
We assume time reversal invariance, 
so as to have a symmetric matrix.
For clarification we note that if 
$N$ particles incident in channel $b$, 
then $g_{ab} N$ particles  
emerge in channel $a$. This means that 
$g_{ab}$ is the ratio between ingoing 
and outgoing fluxes.
In the ergodic state Eq.(\ref{e18})
implies that $\rho_a \propto 1/v_a$.  
Therefore we have detailed balance:  
\be{20}  
g_{ab} \,\rho_b v_b \,\,=\,\, g_{ba} \,\rho_a v_a
\ \ \ \ \ \ \ \mbox{[no summation]} 
\ee 
Namely, for a stationary state the transitions 
from $a$ to $b$ are exactly balanced by the 
transitions from $b$ to $a$.

\section{Definition of the conductance}

If the EMF is not too large we expect to have 
Ohmic behavior ${\langle \mathcal{I} \rangle \propto \mbox{EMF} }$. 
Hence we define the conductance via the relation 
\be{21}
\langle \mathcal{I} \rangle = - G \dot{\Phi}
\ee
This is in fact a special case of a more general 
concept of conductance. In the theory of ``quantum pumping" 
we may vary in time some other parameter $X$, and 
get current  $\langle \mathcal{I} \rangle \propto \dot{X}$. 
Hence we can define (generalized) conductance via   
\be{22}
\langle \mathcal{I} \rangle = -G \dot{X}  
\ee
Much of the formalism that we are going 
to use can be extended to handle this more general case.     
However, in what follows we restrict ourselves  
to the case of an EMF driven system.

Still one can wonder whether it is important 
to specify how the EMF is introduced. 
Does it matter how $a(x)$ look like? 
The answer is that within linear response 
theory all the choices lead to the same 
result. This is not merely a gauge issue 
because for different $a(x)$ 
the electric field along the ring does 
not look the same. Namely, if the gauge 
of the vector potential is changed 
\be{23}
a(x):=a(x)+(d/dx)v(x)
\ee 
then the new scalar potential is 
\be{24}
V(x):=V(x)-e\dot{\Phi} v(x)
\ee
Thus a different choice for $x_0$  
can be regarded as associated with 
adding a rectangular barrier of height $e\dot{\Phi}$. 
Within linear response it is assumed 
that  $e\dot{\Phi}$ is too small 
to make any difference. If this assumption is not applicable, 
it is no longer the ``linear response regime",  
and the specification of $a(x)$ becomes significant.

\section{The long time scenario}

Some people find it inappropriate to define 
conductance for a closed system because 
the problem does not possess a stationary solution \cite{longtime}. 
Namely, it is clear that without a contact 
to a thermal bath the driven system is 
gradually heated up. However, we find this 
objection of no relevance. The practical point 
of view of an electrical engineer 
is demonstrated in Fig.1c. It is clear that 
at any {\em moment} the engineer is inclined 
to characterize the ring by its conductance. 
This is true irrespective of whether 
there is a contact with a thermal bath or not.  
In the absence of such contact it is evident that 
the system is heated up and therefore 
the conductance becomes time-dependent.

It is true that the overall scenario 
is (formally) beyond linear response, 
but it is also true that at a given instant 
of time it is feasible to have a valid linear response 
description. The validity condition is having 
an ergodization time which is much smaller 
than the time that it takes to have a significant 
change in the (evolving) energy distribution 
of the system. This reasoning leads 
to ``slowness conditions'' that are further 
discussed in Ref.\cite{frc}.

We note that there is a strict analogy here 
with the strategy of analyzing quantum pumping. 
Also there the conductance $G(X)$ is calculated, 
using either BPT or the Kubo formula, 
for various points $X=(X_1,X_2)$ in parameter space. 
Later the pumped charge is expressed as 
a line integral over the conductance:
\be{25}
Q = \oint \langle \mathcal{I} \rangle dt  
=  \oint [-G(X(t))\cdot\dot{X}]dt = -\oint G \cdot \vec{dX} 
\ee
Consequently the pumped charge
is not proportional to the amplitude of the 
driving cycle. Thus we can get (globally) 
``non linear response" out of (momentary) 
linear response analysis. 
See \cite{pme} for more details.

\section{Digression: The role of the environment}

On the {\em mathematical} side we have already defined 
precisely the assumptions in the basis of the derivation 
that we are going to present. The purpose of the present 
section is to further clarify the {\em physical} 
circumstances that justify these assumptions. Also 
we would like to put the present work in context of past literature.      
This section can be skipped in first reading.

\subsection{Quantum to classical correspondence}

Our first step is to define a feasibility condition 
for the validity of a semiclassical treatment. 
Assuming that the total transmission of the device  
is $g_T \sim 1$ the time for the randomization 
of the velocity is 
\be{0}
\tau_{cl} \ \approx \ \left(\frac{1}{1-g_T}\right) \times \frac{L}{v_F} 
\ee 
where $L$ is the length of the ring,
and $v_F$ is the Fermi velocity.  
On the other hand the time that it takes 
to resolve the quantized energy levels 
of the ring is 
\be{0}
t_{\tbox{Heisenberg}} \ \approx \ \mathcal{M} \times \frac{L}{v_F}
\ee
Hence the quantum-to-classical correspondence condition is 
\be{0}
\mathcal{M} \ \gg \ \frac{1}{1-g_T}
\ee
which is always satisfied in the classical limit.
Note that the limit $\mathcal{M}\rightarrow\infty$ is 
analogous to $\hbar\rightarrow 0$.

\subsection{Single mode rings}

In the case of a single mode ring 
(a one dimensional ring with a delta scatterer) 
a classical treatment of the dynamics 
does not make any sense in view of the 
correspondence condition of the previous subsection. 
The analysis should be purely quantum mechanical 
and issues such as Landau-Zener transitions \cite{wilk}, 
Debye relaxation mechanism \cite{debye} and  
Dynamical Localization \cite{locl1,locl2} 
should be taken into account. 
Recently we have introduced a new ingredient into the analysis 
that sheds a new light on the whole issue \cite{kbr}.

\subsection{Multi mode diffusive rings}

In the case of multi mode diffusive ring
the leading order result for the conductance 
is as expected just the classical Drude expression. 
The typical calculation \cite{IS} assumes that the 
levels are ``broadened" due to 
the interaction with the environment. 
In the major case of interest 
the level broadening $\Gamma$ is 
assumed to be larger than the mean level 
spacing $\Delta$ but much smaller than 
any semiclassical energy scale. 
Hence it barely affects the Drude result. 
Still it determines the quantum weak localization 
correction, which turns out to be of order $\Delta/\Gamma$.  
The weak localization correction 
depends on the levels statistics and 
therefore on the magnetic flux. Hence it can be 
detected in an actual experiment \cite{orsay}.

\subsection{Multi mode ballistic rings} 
      
In the present work we consider 
a multi mode ballistic ring 
rather than a diffusive ring. 
This means that the time to randomize 
the velocity can be much larger than 
the time $L/v_F$ to make one 
round along the ring. This implies that 
in our configuration the conductance 
(in natural units) can be larger than the 
number of open modes.

We assume that the environment induces
level broadening $\Gamma$ which is larger 
than $\hbar v_F / L$.  
Hence we derive the leading (classical) term 
and do not take into account the implications 
of quantum interference.  
To put this assumption in a larger perspective 
we make the following classification: 
\begin{itemize}
\item[1.]
Isolated system (no environment). 
\item[2.] 
The bath induces only decoherence effect.
\item[3.] 
The bath induces velocity randomization. 
\item[4.] 
Bath limited dynamics.
\end{itemize}
The first case of fully coherent dynamics 
will be analyzed in \cite{kbw}. 
We consider in this work only the 
second case. Still we would like 
to further explain why the second case 
is physically typical, and to make 
some comments on the other two cases.

\subsection{The decoherence mechanism} 

It is typical to assume that the fluctuations 
of the environment are of large spatial correlation length 
compared with $L_{\perp}$. 
An extreme case is the Caldeira-Leggett modeling 
which assumes an infinite correlation length.  
The matrix elements of the position variable 
scale like $L_{\perp}$ for inter-mode transitions 
and like $L$ for intra-mode transitions.
Hence inter-mode transitions are rare compared 
with intra-mode transitions, 
the ratio being $(L_{\perp}/L)^2$. 
Therefore it is realistic to consider 
circumstances such that velocity randomization due 
to the environmental ``noise" can be 
neglected, while intra-mode transitions 
cannot be neglected. The latter lead to decoherence. 
The simplest estimate for the decoherence rate 
is $\Gamma = \eta k_BT L^2/\hbar$,  
where $\eta$ characterizes the coupling to the environmental 
modes, and $T$ is the temperature. 
Thus we see that case~2 is physically typical.

\subsection{Velocity randomization} 

If the fluctuations are strong enough  
inter-mode transitions cannot be ignored. 
This would lead to randomization of the 
velocity in the wire region. 
The scenario of having a bath that 
just randomizes the velocity, but does 
not affect the transmission of the ring 
is apparently not of much physical interest. 
Still such effect can be realized artificially, 
and it is of pedagogical importance, 
as discussed in section~12.

\subsection{Bath limited dynamics} 

If the interaction with the environment 
determines the transmission of the ring,    
we get to case~4.  The most obvious example 
is the the analysis of conductance 
in room temperatures. 
The scattering and hence the diffusion  
of the particles is dominated by bath 
induced inelastic scattering by phonons.

The main point regarding case~4 is 
that the bath cannot be eliminated 
from the model analysis. Another example 
for such circumstances is provided 
by the Debye dissipation mechanism \cite{debye}.
The latter assumes that 
the inelastic relaxation 
time is much shorter compared with 
the adiabatic variation of the energy 
levels, leading to time lag between 
the driving cycle and the adjustment 
of the occupation probabilities.

\section{Linear response analysis}

The simplest route to linear response 
theory \cite{wilk,jar,frc,pmc} takes   
the relation $d \mathcal{H} / dt =\partial \mathcal{H}/ \partial t$
as a starting point. It follows that the change 
in the energy of a particle is given with no 
approximation by the formula 
\be{26}
\mathcal{H}(t) - \mathcal{H}(0)  
= -\dot{\Phi} \int_0^t  \mathcal{F}(t')dt'  
\ee
By squaring and averaging over initial conditions 
we get that the second moment as a double time 
integral over $\langle \mathcal{F}(t') \mathcal{F}(t'') \rangle$. 
Within linear response this correlation is {\em approximated} 
by the {\em stationary} correlation function
\be{27}
C(t'-t'') =  \langle {\cal F}(t') {\cal F}(t'') \rangle_E
\ee 
where the average on the right hand side 
is taken with a zero order microcanonical solution. 
Thus one concludes that there is a diffusion 
in energy, with the coefficient   
\be{28}
D_E = \dot{\Phi}^2  \times \frac{1}{2} \int_{-\infty}^{\infty} C(\tau) d\tau
\ee

Next one wants to see what happens in the 
more general case of an arbitrary~$f(E)$. 
On long times it is argued that the probability 
distribution $\rho(E)$ satisfies 
the following diffusion equation:
\be{29}
\frac{\partial \rho}{\partial t} \ = \
\frac{\partial}{\partial E}
\left(\mathsf{g}(E)D_E \frac{\partial}{\partial E}
\left(\frac{1}{\mathsf{g}(E)}\rho\right)\right)
\ee
The energy of the system is
$\langle {\cal H} \rangle=\int E \rho(E)dE$.
It follows that the rate of energy absorption is
\be{30}
\frac{d}{dt}\langle {\cal H} \rangle
= - \int_0^{\infty} dE \ \mathsf{g}(E) \ D_E
\ \frac{\partial}{\partial E}
\left(\frac{\rho(E)}{\mathsf{g}(E)}\right)
\ee
For zero temperature Fermi occupation we get
\be{31}
\frac{d}{dt}\langle {\cal H} \rangle
\ = \
\left[ \mathsf{g}(E) \frac{}{} D_{E}  \right]_{E=E_F} 
\ = \ G \dot{\Phi}^2
\ee
This is the mesoscopic version of Joule law. 
The expression for the conductance is 
\be{32}
G = \mathsf{g}(E_F) \times 
\frac{1}{2}\int_{-\infty}^{\infty} C(\tau)d\tau
\ee

\section{The Kubo formula}

The above is apparently the simplest and most illuminating 
derivation of the Fluctuation-Dissipation version 
of the Kubo formula. A more complicated treatment \cite{pmc,pmo,pme} 
allows to write a generalized version that holds also 
for ``quantum pumping" applications. Namely, 
\be{33}
G = \mathsf{g}(E_F) \times 
\int_{0}^{\infty} C(\tau) d\tau
\ee
with 
\be{34}
C(\tau) =  \langle {\cal I}(\tau) {\cal F}(0) \rangle_E
\ee 
where $\mathcal{F}$ is the generalized force 
which is associated with the driving. 
In the present application $\mathcal{F}$ is 
just the current operator, hence $C(\tau)$ 
is symmetric, and therefore the generalized 
version Eq.(\ref{e33}) is equivalent to Eq.(\ref{e32}).

It is also important to make a connection 
with the more traditional treatment of conductance 
in case of disordered metals.
If we set $a(x)=1/L$ for the vector potential 
we get $(e/L)\hat{v}$ as the current operator. 
Hence we get from Eq.(\ref{e32})
\be{35}
G = \mathsf{g}(E_F) \times 
\frac{1}{2}
\left(\frac{e}{L}\right)^2
\int_{-\infty}^{\infty} \langle v(\tau) v(0) \rangle d\tau
\ee
The conventional derivation of the Drude 
formula is based on the assumption of exponential 
decay of the velocity-velocity correlation function. 
The latter formula is (formally) valid in any case, 
but in case of a disordered sample it has special 
appeal because it implies the Einstein relation between 
the conductance and the spatial diffusion. 
Namely, if we have a diffusive ring then 
it is natural to write $G=\hat{G}/L$ 
and to define $\hat{\mathsf{g}}(E_F)=\mathsf{g}(E_F)/L$. 
Then we can rewrite Eq.(\ref{e35}) as 
\be{36}
\hat{G} = e^2 \hat{\mathsf{g}}(E_F) D_{\tbox{space}}
\ee 
The more conventional derivations of this expression 
is based on the phenomenological relation 
$J = -D \nabla (\mbox{density}) - \hat{G} \nabla (\mbox{potential})$ 
and the argument that $J=0$ at equilibrium.

\section{Implications of the Kubo formula}

From the the Kubo formula Eq.(\ref{e33})
it is not obvious that the result for $G$ is 
independent of where we measure $\mathcal{I}$. 
In general it can be proved \cite{shift} 
that for a different choice of $\mathcal{I}$ 
the corresponding Kubo conductance may 
differ at most by  
$e\mathsf{g}(E_F)\langle\mathcal{F}\rangle$. 
But if $\mathcal{F}$ is a current operator 
then $\langle\mathcal{F}\rangle=0$ and therefore 
the conductance be becomes independent of $x_1$.

Also it seems that the Kubo conductance 
is proportional to the density of states. 
Therefore, if we had doubled the volume 
of the cavity, would we get larger conductance?
Furthermore, does the result for $G$ depends 
merely on the transition matrix, 
and not (say) on the dwell time inside 
the scattering region? 

To answer these questions, and to establish 
the $x_1$ independence of $G$,  let us write 
the Kubo formula in a more illuminating way. 
By definition we have 
\be{37}
&& \int_{-\infty}^{\infty} 
\langle {\cal I}(\tau) {\cal F}(0) \rangle  d\tau
= \sum_r p_r {\cal F}_r Q_r 
\nonumber \\ 
&& \equiv \frac{1}{\mathsf{g}(E)} \int  \delta(E-\mathcal{H}(\bm{r},\bm{p})) 
\frac{d\bm{r}d\bm{p}}{(2\pi)^d} F(\bm{r},\bm{p})Q(\bm{r},\bm{p})
\ee
where $r$ is an index that labels phase space 
cells (different initial conditions), and $p_r$ 
corresponds to a microcanonical distribution. 
We have introduced the notation
\be{38}
Q(\bm{r},\bm{p}) = 
\int_{-\infty}^{\infty} 
{\cal I}(\tau;\bm{r},\bm{p}) d\tau
\ee
Namely $Q(\bm{r},\bm{p})$ is the total charge 
which is obtained by integrating the current 
which is induced by a particle that goes through 
the point $(\bm{r},\bm{p})$ at $t=0$. 
It is in fact (for $e=1$) the winding number of 
the associated trajectory, and therefore it gives 
a result which is independent of the chosen section. 
Note however that $Q(\bm{r},\bm{p})$ obtains 
a meaningful value only upon course graining, 
else it is erratic. 
Now we can write the Kubo formula as 
\be{39}
G = \frac{1}{2} \int  \delta(E-\mathcal{H}(\bm{r},\bm{p})) 
\frac{d\bm{r}d\bm{p}}{(2\pi)^d} F(\bm{r},\bm{p})Q(\bm{r},\bm{p})
\ee
This expression has several advantages. 
One advantage we have mentioned: the result 
is manifestly independent of the 
definition of the current operator. 
The second advantage is that it shows that 
the global density of states is in fact 
not important. We can double the volume 
of the cavity, still we would get the same 
result provided that the scattering probabilities 
are not affected. In particular we see that 
time delays are not important.

\section{Conductance of a single mode ring (part 1)}

In this section we show how the Kubo formula 
for a {\em closed} ring leads to a Landauer-alike formula 
for the conductance provided the 
effect of the environment is to completely 
randomize the velocity within the wire region
without affecting its transmission.
To simplify the presentation we consider the single mode case.

For the Kubo formula we have to evaluate 
the correlation function $C(\tau)$ of Eq.(\ref{e34}), 
and to calculate the integral in Eq.(\ref{e33}).
As explained the result of the calculation 
should be independent of the how we define 
the current operators. The simplest choice is to 
define $\mathcal{F}$ as the current through a section 
$x=x_0$ on the left of the scattering region, 
while $\mathcal{I}$ is the  current through a section 
$x=x_1$ on the right of the scattering region.
$C(\tau)$ comes out as a sum of delta functions. 
The shortest time correlation is associated 
with the time $\tau_1$ to cross the scattering 
region. For example, if there is no time delay 
then $\tau_1=(x_1-x_0)/v_E$. 
We can regularize $\mathcal{F}$ 
as a rectangular of width $\varepsilon$. 
The probability to have there a particle 
moving in the right direction, 
such that $\mathcal{F}(0)=ev_E/\varepsilon$  
is $(\varepsilon/L)/2$. The current 
that we get in the other side of the barrier 
is $\mathcal{I}=e\delta(\tau-\tau_1)$. 
Assuming that this is the only correlation, 
and taking into account the time reversed 
correlation for $\tau<0$, we get
\be{40}
C(\tau) =
e^2\frac{v_{\tbox{F}}}{L} 
\sum_{\pm} \frac{1}{2}g_T \delta(\tau \mp \tau_1)
\ee
Note that if we had chosen $x_1=x_0$ 
we would get three delta functions: 
a self correlation delta function $\delta(\tau)$ 
and reflection peaks. Namely,  
\be{41}
C(\tau) =
e^2\frac{v_{\tbox{F}}}{L}  
\left[ \delta(\tau) - \frac{1}{2}(1-g_T) \sum_{\pm} \delta(\tau\mp\tau_0) \right]
\ee
where $\tau_0$ is the scattering time.
Obviously the integral over the new $C(\tau)$ 
is the same as the integral over the former one.
Irrespective of our choices we get from Kubo  
\be{42}
\bm{G} =  
\frac{e^2}{2\pi\hbar} g_T
\ee
which looks like the (single mode) Landauer formula.

\section{Conductance of a single mode ring (part 2)}

If the velocity is not randomized 
within the wire, then there are other 
correlations that involve the time $L/v_E$ 
to encircle the ring. 
For the following calculation it is simplest 
to assume that $0<x_0<x_1$. What we have is 
to calculate the integral 
\be{43}
\int_0^{\infty} 
\langle {\cal I}(\tau) {\cal F}(0) \rangle  d\tau
= \sum_r p_r {\cal F}_r Q_r 
\ee
Note that we find it convenient here  
to set $\tau=0$ as the lower bound 
of the integral. Recall that $r$ is an 
index that labels phase space 
cells (different initial conditions), 
and $p_r$ corresponds to a microcanonical 
distribution. 
Note that ${\cal F}$ is non-zero 
only if $r$ is located at $x=x_0$.
The total charge which is transported through 
the section $x=x_1$ is defined here as 
\be{44}
Q_r = \int_0^{\infty} \langle{\cal I}(\tau)\rangle_r  d\tau
\ee 
where the current is evaluated under 
the assumption that the particle is launched 
at point~$r$. There are two relevant possibilities:  
Either the particle is launched at $x=x_0$ 
in the clockwise direction, 
or it is launched at $x=x_0$ 
in the anti-clockwise direction. 
Observe that the (net) charge that  
goes through the section $x=x_1$ 
after a round trip is suppressed 
by a factor $(2g_T-1)$ 
due to the scattering (we sum the clockwise 
and the anticlockwise contributions).
Thus we get that the total charge that 
goes through the section is 
\be{45}
Q^{\rightarrow} 
=  e\left[ 1 +  (2g_T-1) + (2g_T-1)^2 + ... \right]
=  e\left[\frac{1}{2(1-g_T)}\right]
\ee
for a particle that is launched clockwise, 
and 
\be{46}
Q^{\leftarrow} = - e\left[ \frac{1}{2(1-g_T)} - 1 \right]
\ee
for a particle that is launched anti-clockwise
Thus we get
\be{47}
\int_0^{\infty} 
\langle {\cal I}(\tau) {\cal F}(0) \rangle  d\tau 
= \frac{1}{2L} (+ev_{\tbox{E}}) Q^{\rightarrow} + \frac{1}{2L} (-ev_{\tbox{E}}) Q^{\leftarrow}
= e^2 \frac{v_{\tbox{E}}}{L} \left[\frac{g_T}{1-g_T}\right]
\ee
leading to 
\be{48}
\bm{G} =  
\frac{e^2}{2\pi\hbar} \left[ \frac{g_T}{1-g_T} \right] 
\ee

\section{Relation to the Landauer formula}

As explained in section~10 we get from the Kubo formula 
a Landauer look-alike formula if we assume 
that the environment induces velocity 
randomization in the wire region without 
affecting its transmission.   
In fact we can get to the same conclusion by modeling 
the ``loss of memory'' in the wire region as 
a scatterer with transmission $g_{\tbox{wire}}=1/2$. 
It is well know that the $G$ of Eq.(\ref{e48}) 
obeys Ohm law for addition of resistors in series. Hence 
\be{49}
\bm{G} =  
\frac{e^2}{2\pi\hbar}  
\left( \frac{g_T}{1{-}g_T} \right) 
= \frac{e^2}{2\pi\hbar}  
\left[ 
\left( \frac{g_0}{1{-}g_0} \right)^{-1} 
+\left( \frac{g_{\tbox{wire}}}{1{-}g_{\tbox{wire}}} \right)^{-1} 
\right]^{-1} 
=
\frac{e^2}{2\pi\hbar} g_0
\ee

We would like to emphasize that the purpose of this 
section is purely pedagogical. As stated in 
section~6 an environment that just randomize 
the velocity without affecting the transmission         
is apparently of no physical interest.  
Still if one insists it can be constructed 
artificially. Simply cut the wire and connect 
the two ends to a chaotic cavity. A particle 
that moves in the wire gets into the cavity and 
after a time delay gets out either 
clockwise or anti-clockwise with equal 
probabilities. Hence in such arrangement $g_{\tbox{wire}}=1/2$.

The pedagogical importance of the above discussion 
is in making a bridge between the reservoir 
philosophy of the Landauer construction and the 
Kubo formalism of closed systems.  The memory loss 
device that we have described above provides 
the same ``service" as the reservoirs in the Landauer picture.

\section{Conductance of a multi-mode ring}

Let us assume that the EMF  $-\dot{\Phi}$ is concentrated 
across the scattering region ($x_0=0$).
When a particle goes through $x=0$ it gains momentum  
($p \mapsto p-e\dot{\Phi}/v$), 
where $v=|p|/\mathsf{m}$. 
Hence the change is energy is $E\mapsto E\mp e\dot{\Phi}$ 
for right and left movers respectively.  
The state of the system is described 
by the distribution functions $\rho_n^{\rightarrow}(E)$ 
and $\rho_n^{\leftarrow}(E)$ of Eq.(\ref{e17}). 
The index $n$ distinguishes different modes. 
It is implicit from now on that we look for 
an ergodic-like solution, such that the density of 
the particles along the ring is uniform.
The balance equations are:    
\be{50}
\frac{\partial\rho_n^{\rightarrow}}{\partial t} 
&=& 
- \left[\rho_n^{\rightarrow}v_n\right]  
+ \left[ \sum_m  g^T_{nm} \rho_m^{\rightarrow}v_m\right]_{E+e\dot{\Phi}}
+ \left[ \sum_m  g^R_{nm} \rho_m^{\leftarrow}v_m\right]  \\
\frac{\partial\rho_n^{\leftarrow}}{\partial t} 
&=& 
- \left[\rho_n^{\leftarrow}v_n\right]  
+ \left[ \sum_m  g^T_{nm} \rho_m^{\leftarrow}v_m\right]_{E-e\dot{\Phi}}
+ \left[ \sum_m  g^R_{nm} \rho_m^{\rightarrow}v_m\right] 
\ee
It can be verified that the zero order ($\dot{\Phi}=0$) 
stationary solution of this equation is given by Eq.(\ref{e18}),  
where $f(E)$ is an arbitrary function. 
We are looking for a first order stationary-like solution. 
The linearized equation for the clockwise moving particles is  
\be{52}
\left[ \sum_m  (1-g^T)_{nm} \delta \rho_m^{\rightarrow}v_m\right] 
- \left[ \sum_m  g^R_{nm} \delta \rho_m^{\leftarrow}v_m\right] 
=  
\frac{e}{2\pi}\dot{\Phi} \frac{\partial f(E)}{\partial E}  \sum_m g^T_{nm}
\ee
A similar equation exist for the counter-clockwise particles. 
Subtracting the corresponding equations we get 
\be{53}
\sum_m [(1-g^T)+g^R]_{nm}[\rho^{\rightarrow}v-\rho^{\leftarrow}v]_m 
= 2\frac{e}{2\pi}\dot{\Phi} \frac{\partial f(E)}{\partial E} g_{n}
\ee
with the solution
\be{54}
[\rho^{\rightarrow}v-\rho^{\leftarrow}v]_n =
\frac{e}{2\pi}\dot{\Phi} \frac{\partial f(E)}{\partial E}
\sum_{n'} \left[\frac{2}{(1-g^T)+g^R}\right]_{nn'} g_{n'}
\ee
The current is 
\be{55}
\mathcal{I} &=& \sum_n \int_0^{\infty} dE 
\ (\rho_n^{\rightarrow}-\rho_n^{\leftarrow}) ev_n  
\nonumber \\
&=&  \dot{\Phi}  \frac{e^2}{2\pi}
\int_0^{\infty} 
\sum_{nm} \left[ \left(\frac{2}{1-g^T+g^R}\right)g^T \right]_{nm} 
\frac{\partial f(E)}{\partial E} dE 
\ee 
With the assumption of Fermi occupation we get Eq.(\ref{e3}).
Note that upon summation the order of matrix multiplication 
is not important because $g_{nm}$ is symmetric.

\section{The wire with cavity model system}

We consider a ring (Fig.1b) which is formed by 
folding a rectangular waveguide 
(i.e. imposing periodic boundary conditions). 
A chaotic cavity is attached to the 
waveguide at one ``point". A particle 
has some probability to enter the cavity, 
where memory is ``lost", and then it gets 
out again with equal probability 
to either side. 
A particle that travels in mode $n$ 
of the waveguide has a transverse 
momentum $\pm(\pi\hbar/L_{\perp})n$ 
where $L_{\perp}$ is the width of the waveguide.
The distance between subsequent hits 
of the same wall is    
\be{56} 
\mbox{step} = 
\frac{ \sqrt{2\mathsf{m}E-((\pi\hbar/L_{\perp})n)^2} }
{ (\pi\hbar/L_{\perp})n }  
2L_{\perp}
= \frac{\sqrt{\mathcal{M}^2-n^2}}{n} 2L_{\perp}
\ee
The number of open modes $\mathcal{M}$ 
is implicitly defined via the latter equality.
The probability to get into the cavity 
via an opening of size $L_{\tbox{op}}$ is:
\be{57} 
p_n = 
\frac{L_{\tbox{op}}}{\mbox{step}} 
= \mbox{minimum}
\left[
\frac{ \alpha }{ \sqrt{(\mathcal{M}/n)^2-1} },\,\, 1 
\right]
\ee  
where $\alpha=L_{\tbox{op}}/(2L_{\perp})$. 
The crossover from $p_n<1$ to $p_n=1$ happens at
\be{58}
n_c = \frac{1}{\sqrt{1+\alpha^2}} \mathcal{M}
\ee 
We are going to treat $\mathcal{M}$ as a free 
parameter. Hence we have two parameters 
that characterize the scattering: the 
classical (geometrical) parameter $\alpha$, 
and the quantum-mechanical parameter 
$\mathcal{M}$. Note that the classical 
limit is $\mathcal{M}\rightarrow\infty$.

Let $q_n$ be the probability 
to get out of the box to mode $n$, 
either to the right going channel 
or to the left going channel.  
It follows that $g^R_{nm} = (1/2) q_n p_m$.
From $g_{nm} = g_{mn}$ we conclude  
that $q_n/p_n=c$ is the same for all channels. 
Taking into account that $\sum_n q_n = 1$ 
we get $c=1/(\sum_n p_n)$, and hence  
\be{59} 
g^R_{nm} &=& \frac{1}{2} q_n p_m  
= \frac{c}{2} p_n p_m   
\\ \label{e60}
g^T_{nm} &=& \frac{1}{2} q_n p_m + (1-p_m)\delta_{nm}  
= \frac{c}{2} p_n p_m + (1-p_m)\delta_{nm}  
\ee
Thus, given the input 
parameters $\alpha$ and $\mathcal{M}$, 
we can calculate $g_{nm}$. 
It is useful to define the total 
probability of transmission for 
a particle that comes in channel $n$ as: 
\be{61}
g_n &\equiv& \sum_m g^T_{mn} = 1 - \frac{1}{2}p_n
\ee  
For sake of later estimates we note that 
for $\mathcal{M}\gg1$, 
sums over $n$ can be approximated 
by an integral over $x=n/\mathcal{M}$. Using  
the obvious notation $x_c=n_c/\mathcal{M}$ we get 
\be{62}
\frac{1}{\mathcal{M}}\sum p_n  &\approx&
\alpha \int_0^{x_c} [(1/x)^2-1]^{-1/2} dx + [1-x_c] 
\nonumber \\ &=&
\alpha \frac{1}{2} \left[ 1- (1-x_c^2)^{1/2} \right] 
+ [1-x_c]
\nonumber \\ &=& 
1+\frac{1}{2}\alpha
-(1+\alpha^2)^{-1/2}\left(1+\frac{1}{2}\alpha^2\right)
\nonumber \\
&\approx& \frac{1}{2}\alpha +\mathcal{O}(\alpha^4)
\ee 
and
\be{63}
\frac{1}{\mathcal{M}}\sum \frac{1}{p_n}  &\approx&
\frac{1}{\alpha}\int_{1/\mathcal{M}}^{x_c} [(1/x)^2-1]^{1/2} dx + [1-x_c] 
\nonumber \\ &=& 
\frac{1}{\alpha}
\left[ 
\ln\left( \frac{2x_c}{1+(1-x_c^2)^{1/2}} \mathcal{M} \right)         
-\left( 1 - (1-x_c^2)^{1/2} \right) 
\right] 
+ [1-x_c]
\nonumber \\ &=&
\frac{1}{\alpha}
\left[ 
\ln\left( \frac{2}{\alpha+(1+\alpha^2)^{1/2}} \mathcal{M} \right)         
-1
\right] 
+1
\nonumber \\ 
&\approx&
[\ln(2\mathcal{M})-1]\frac{1}{\alpha} + \mathcal{O}(\alpha)
\ee

\begin{figure}[b]
\putgrf[0.3\hsize]{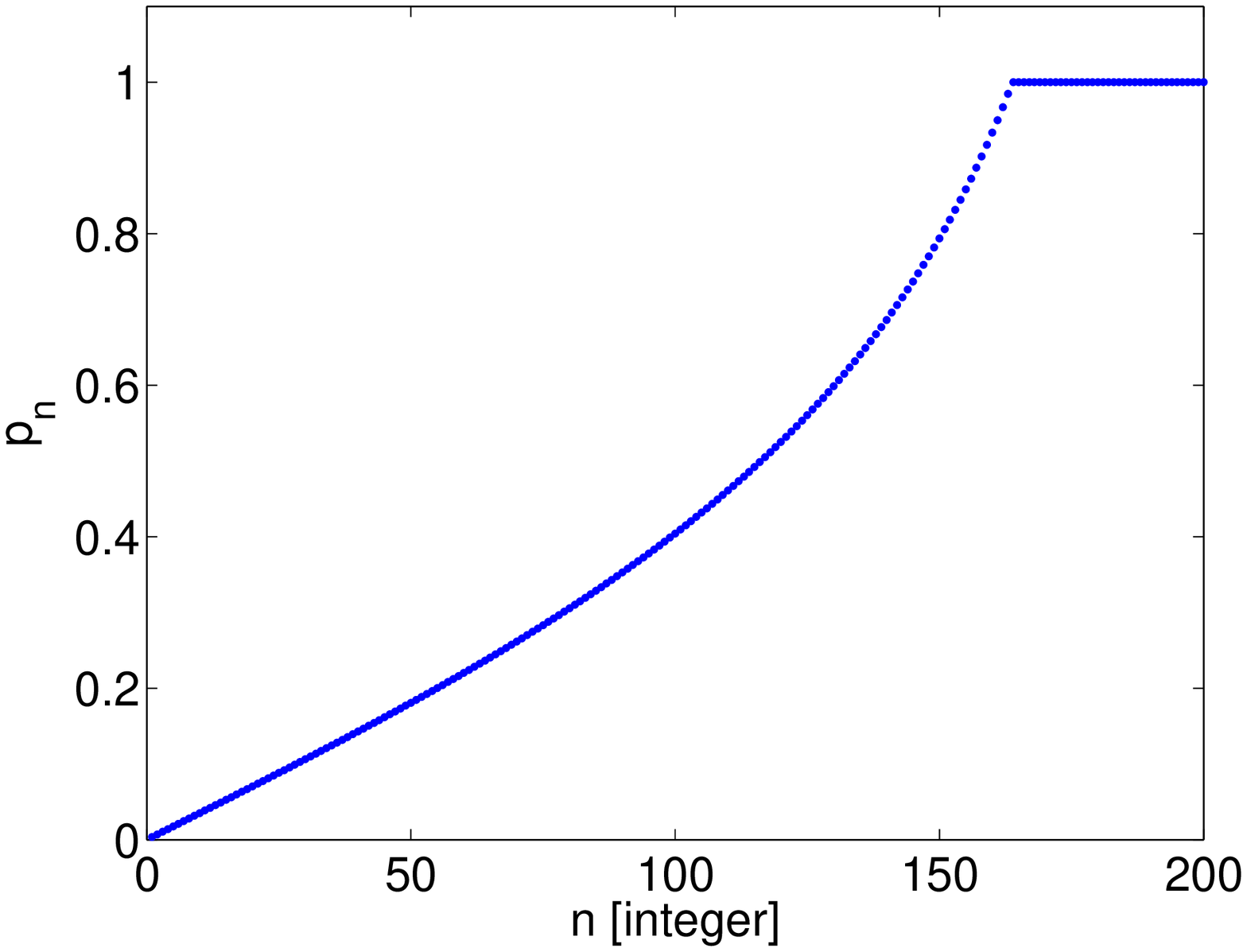}
\putgrf[0.3\hsize]{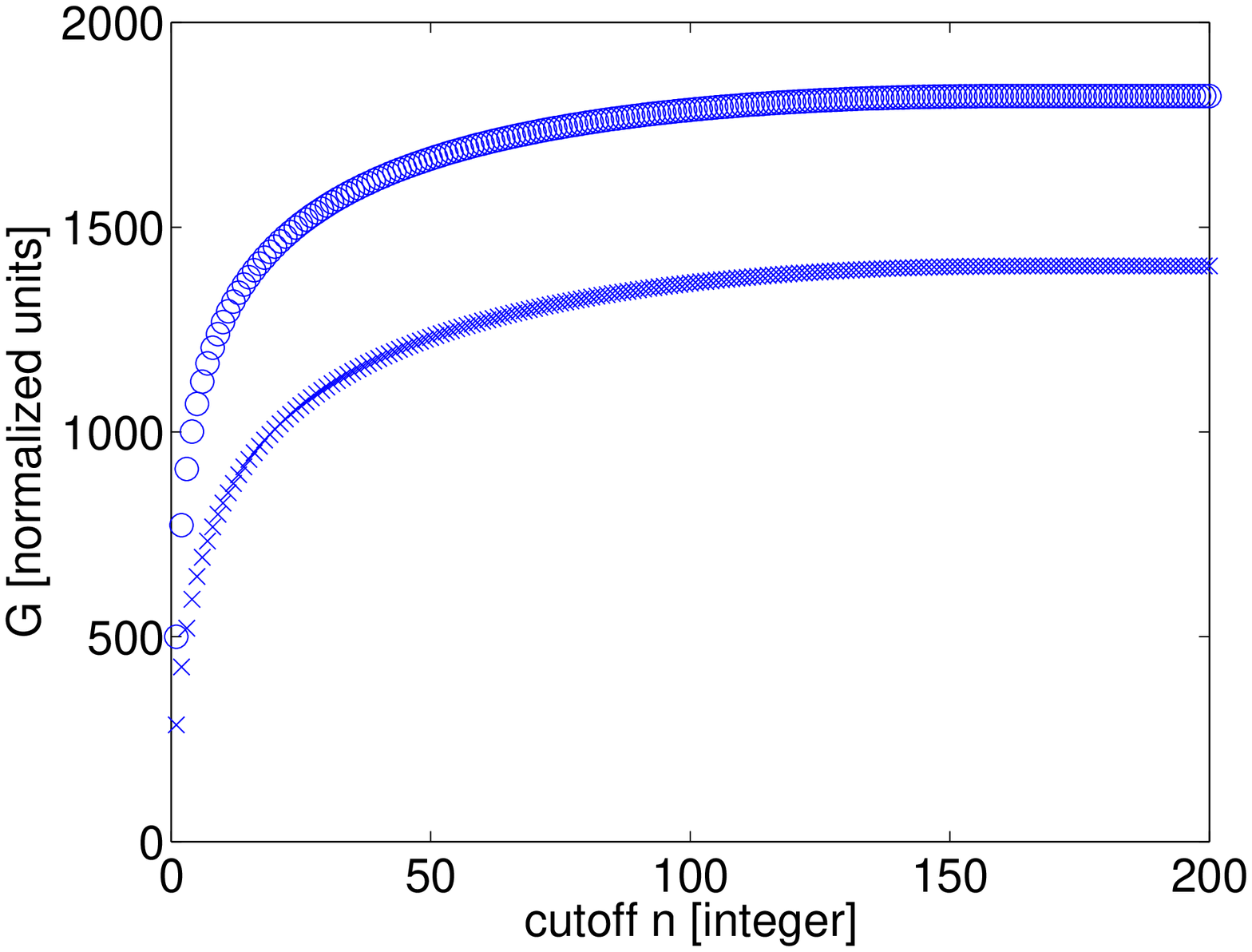}
\caption{The left panel displays $p_n$ 
for a system with $\alpha=0.7$ and $\mathcal{M}=200$ open modes. 
The crosses in the right panel are for the cumulative sum 
over $(1-p_n)/p_n$.  The circles are calculated from the 
exact formula. Namely, the matrix $h$ is diagonalized, 
and the cumulative trace over the square of its elements 
is displayed.}
\end{figure}

We now turn to the calculation of the conductance. 
First of all, let us calculate the 
Landauer conductance. Thanks to the simple 
structure of the $g^T_{nm}$ matrix, the 
calculation is quite easy
\be{64} 
G_{\tbox{Landauer}} 
= \frac{e^2}{2\pi\hbar} \sum_{n,m} g^T_{nm}
= \frac{e^2}{2\pi\hbar} \sum_n g_n
\ee
Each channel has total transmission 
in the range $(1/2)<g_n<1$ and therefore  
the conductance (in normalized units) 
roughly equals to the number of open modes.
Substitution of (\ref{e61}) leads to:
\be{65}
G_{\tbox{Landauer}} = \frac{e^2}{2\pi\hbar}\sum_n (1-(p_n/2)) 
=  \frac{e^2}{2\pi\hbar}[1 - \frac{1}{4}\alpha +\mathcal{O}(\alpha^4) ] \mathcal{M}
\ee

For the multimode conductance of Eq.(\ref{e3}) 
the calculation is more complicated.
At first sight it seems that the calculation 
should be done numerically as in Fig.~2. 
The numerical calculation in Fig.~2 (circles) is done 
in a way which is inspired by a similar 
type of calculation within the Landauer formalism.
We define $h_{nm} = ([ 2g^T/(1-g^T+g^R) ]_{nm})^{1/2}$ 
and write the sum in Eq.(\ref{e3}) as $\trc\left[ h^{\dag} h \right]$. 
Then we make singular value decomposition 
of $h$ and sum over the squares of its eigenvalues.

The other way to calculate the multimode conductance 
starts with an attempt to make a zero order 
evaluation of the sum. This means setting $c=0$ in Eq.(\ref{e60}). 
The resulting estimate gives a rough approximation 
as seen from Fig.2 (crosses). The main source 
of error are evidently the low modes.
Surprisingly it turns out that 
the calculation can be carried out to 
infinite order in $c$, thanks to miraculous 
cancellations. Using the expansion 
\be{66}
\frac{1}{A-cB} = \frac{1}{A}
+c\frac{1}{A}B\frac{1}{A}
+c^2\frac{1}{A}B\frac{1}{A}B\frac{1}{A}+... 
\ee
with $A_{nm}=p_n\delta_{nm}$ and $B_{nm}=(1/2)p_np_m$, 
we get the result 
\be{67}
\left[\frac{1}{1-g^T}\right]_{nm} &=& 
\frac{1}{p_n}\delta_{nm} 
+ (c/2) \frac{1}{p_n}p_np_m \frac{1}{p_m} +
\nonumber \\
&+& (c/2)^2 \sum_k \frac{1}{p_n}p_np_k \frac{1}{p_k} p_k p_m \frac{1}{p_m}
+ ... = \frac{1}{p_n}\delta_{nm} + c 
\ee
where in the last step we have made 
a geometric summation over all orders.
Now we can calculate the conductance 
\be{68} 
G =
\frac{e^2}{2\pi\hbar} \sum_{n,m} \left[\frac{g^T}{1-g_T}\right]_{nm}
=
\frac{e^2}{2\pi\hbar}
\left[
\left(\sum_n \frac{1}{p_n}\right) 
+ c \mathcal{M}^2 - \mathcal{M}
\right]
\ee
Recall that $c=1/\sum_n p_n$. Hence this expression 
requires merely the evaluation of the sums 
$\sum_n p_n$ and  $\sum_n (1/p_n)$. 
If we have $p_n \approx 1$ for all modes, 
then we get simply $G = (e^2 / (2\pi\hbar)) \mathcal{M}$ 
which reflects that number of open modes. 
But the interesting case is when $\alpha$ is small: 
\be{69}
G \approx \frac{e^2}{2\pi\hbar}
\left[
\frac{1}{\alpha} (1 + \ln(2\mathcal{M})) 
-1 + \mathcal{O}(\alpha) 
\right] 
\mathcal{M}
\ee
Unlike the case of the Landauer conductance, the 
result does not reflect the number of open modes. 
The contribution of the low modes is singular 
in the limit of small $\alpha$. Furthermore, 
the conductivity (conductance per channel) diverges 
logarithmically in the classical limit.

\section{Concluding remarks}

Much of the derivations in this paper 
can be generalized to analyze ``quantum pumping". 
In Ref.\cite{pmt} we have considered  
a single mode device where the  
current is induced by translating 
a scatterer. Namely, $\mathcal{I}=-G\dot{X}$
and hence the transported charge is $dQ=-GdX$ 
where $dX$ is the displacement of the scatterer. 
From the BPT formula one obtains 
\be{70} 
G \ = - (1-g_0) \times \frac{e}{\pi}k_{\tbox{F}}
\ee
where $g_0$ is the transmission 
of the scatterer and $k_{\tbox{F}}$ 
is the Fermi momentum. If we close 
the system into a ring, and use 
the same assumptions as in this 
paper we get 
\be{71}
G = -
\left[ \frac{1-g_0}{g_0}\right]
\left[ \frac{g_T}{1-g_T}\right]
\times 
\frac{e}{\pi}k_{\tbox{F}} 
\ee
where $g_T$ is the overall 
transmission of the device. 
Multi-mode generalizations of these 
results can be obtained by 
employing either 
the Kubo or the Master equation approach 
as in the present paper (not published).

We regard the ``Ohmic" problem that has been 
discussed in the present paper, and 
the above mentioned ``pumping" problem 
of Ref.\cite{pmt}, as the prototype models 
for the application of linear response 
theory: The ``Ohmic" problem has to do 
with the dissipative part of the response, 
while the ``pumping" problem has to do 
with the geometric (non-dissipative) part 
of the conductance matrix. In both cases 
we can use the Kubo formalism as a starting point, 
and in both cases we can regard the 
scattering approach (Landauer-BPT)
as a special limiting case. 
However, one should be aware of the subtle 
differences between the two problems.  
The main point to remember is that 
the adiabatic limit of ``quantum pumping"  
is the non-vanishing ``adiabatic transport" formalism, 
while the adiabatic limit of 
the ``Ohmic" problem gives zero conductance. 
For further details see \cite{pme}.

We have assumed that the coherence time 
is short compared with the time that it takes 
to encircle the ring. This makes the calculation 
insensitive to the energy~$E$. 
Once we consider a strictly isolated system, 
we get an $E$~dependence that has to do with 
semiclassical energy scales such as $(\hbar/L)v_F$.
Such energy scales are larger compared with 
the mean level spacing ($\propto\hbar^d$), 
and may invalidate the Kubo formalism. 
An extreme example for the implication 
of having a non-universal energy scale is 
analyzed in in Ref.\cite{kbr}. The study of the 
general multi-mode case introduces further 
conceptual as well as technical complications \cite{kbw}.


\ \\

\ack

D.C. has the pleasure to thank Michael Wilkinson  
Bernhard Mehlig and Yuval Gefen for intriguing discussions.  
Many ideas were inspired by the on going collaboration  
with Tsampikos Kottos and Holger Schanz.   
We also thank Renaud Leturcq, Yuval Oreg and Boris Shapiro 
for helpful conversations.  
The Benasque center for science is acknowledged 
for providing the atmosphere for some of these interactions.       
The research was supported by the Israel Science Foundation (grant No.11/02),
and by a grant from the GIF, the German-Israeli Foundation 
for Scientific Research and Development.


\end{document}